\documentclass[a4paper]{article}
\usepackage{RRA4}
\usepackage{times}
\usepackage[latin1]{inputenc}
\usepackage{amsmath}
\usepackage{graphicx}
\usepackage{amssymb}

\newtheorem{thm}{Theorem}
\newtheorem{cor}{Corollary}

\newcommand{\mc}[1]{\mathcal{#1}}

\newcommand{\mM}{\mathcal{M}}

\newenvironment{Proof}{\noindent \textbf{Proof:
}}{\hspace{\stretch{1}}$\square$}

\usepackage[pdfpagelabels,hyperfootnotes=false]{hyperref}

\RRdate{December 2006}
\RRauthor{Dmitry Lebedev\thanks[FT]{France Telecom R\&D, 38--40, rue du g\'{e}n\'{e}ral Leclerc, 92130 Issy les Moulineaux, France} \and Fabien Mathieu\thanksref{FT} \and Laurent Viennot\thanks[INRIA]{INRIA, domaine de Voluceaux, 78153 Le Chesnay cedex, France} \and Anh-Tuan Gai\thanksref{INRIA} \and Julien Reynier\thanks{LIENS, 45 rue d'Ulm, 75230 Paris Cedex 05, France} \and Fabien de Montgolfier\thanks{LIAFA, 175, rue du Chevaleret, 75013 Paris France}}%%
\RRtitle{Applications de la théorie des affectations à l'étude des réseaux P2P}
\RRetitle{On Using Matching Theory to Understand\\ P2P Network Design}

\titlehead{On Using Matching Theory to Understand P2P Network Design}

\RRnote{CRC MARDI}
\RRabstract{
This paper aims to provide insight into stability of collaboration choices in P2P networks.
We study networks where exchanges between nodes are driven by the desire to receive the best service available. This is the case for most existing P2P networks. We explore an evolution model derived from \emph{stable roommates} theory that accounts for heterogeneity between nodes.
We show that most P2P applications can be modeled using stable matching theory. This is the case whenever preference lists can be deduced from the exchange policy. In many cases, the preferences lists are characterized by an interesting \emph{acyclic} property.
We show that P2P networks with acyclic preferences possess a unique stable state with good convergence properties.
}

\RRkeyword{P2P networks, stable matchings, convergence properties}

\RRresume{Cet article vise à décrire la possible stabilité d'un réseau pair-à-pair en termes de collaborations. Nous étudions les réseaux où les échanges entre pairs sont basés sur la volonté d'obtenir le meilleur service possible (hypothèse valable pour une grande partie des réseaux P2P existants). Nous considérons un modèle d'évolution basé sur la théorie des \emph{stable roommates} afin de prendre en compte l'hétérogénéité des pairs. Nous montrons comment étudier la plupart des réseaux P2P à l'aide de ce modèle. Dans la plupart des cas rencontrés, le système de préférences sous-jacent possède une propriété d'acyclicité. Pour de tels systèmes, nous montrons l'existence d'une unique solution stable avec de bonnes propriétés de convergence.}

\RRmotcle{pair-à-pair, mariages stables, propriétés de convergence}%%
\RRprojets{Gyroweb}
\RRtheme{\THCom}% \THCog \THSym \THNum \THBio} % cas de 5 themes
\URRocq % pour ceux qui sont au Sud.

\begin{document}

\makeRR

%\vspace{-6mm}
\section{Introduction}
%\vspace{-4mm}

During the last few years, peer-to-peer (P2P) applications have emerged and
%haved imposed themselves 
now appear as a major component of the Internet, from both
traffic and content distribution points of view. Peer-to-peer network
performances scale reasonably with the number of users. This definitely
makes P2P a leading paradigm for tomorrow's networking applications. One of the most 
striking achievements  is certainly BitTorrent  protocol~ \cite{cohen03incentives} for content distribution. It is based on a Tit-for-Tat mechanism, which is used to regularly compute collaboration links between the peers. Despite the simplicity of this approach, measurements and analysis are hard to provide due to the massively distributed nature of the applications. Starting from this statement, we present a 
formal method to analyze a large class of networks (including BitTorrent-like applications) 
with regard to the stability of collaborations. 

More precisely, our main contribution is a model that fits any peer-to-peer
protocol where peers are allowed to choose partners they are collaborating with. We just suppose that each peer ranks other peers according to some preference function. 
For example, in a P2P file-sharing network, each peer can rank other peers 
according to  the similarity of their interest.
In a cooperative download application such as BitTorrent \cite{cohen03incentives},
the upload bandwidth appears as a major parameter as the incentive mechanism consists
in selecting collaborators based on how much they upload.
Additional parameters like download bandwidth, latency, storage capacity or even manual choices can also be used to define the preferences. 
Even though exact mechanisms of P2P solutions can be more complicated,
our modeling gives a first approximation. In particular, it explains
the evolution of a system where collaborators are selected according to such a parameter.

This work aims to deduce properties of the connection graph induced by the preference system chosen by a P2P application. The paper is structured as follows. In the next section, we present existing formal theories, namely the stable marriage problem~\cite{gale62college} and 
the stable roommate problem~\cite{fleiner05generalization}.
In Section~\ref{sec:model} we apply these theories to develop a model 
that includes many P2P applications. 
Section~\ref{sec:acyclic} identifies three main preference classes that appear in
existing P2P applications. All of them appear to be cycle free and
can be seen as particular cases of what we call acyclic preferences.
In Section~\ref{sec:stability} we  focus on acyclic preference systems and present a corresponding stability result. The convergence speed is discussed in Section~\ref{sec:convergence}. 

\section{Background: matching theory}
\label{sec:background}

Stable marriage problems were introduced by Gale and Shapley in 1962
\cite{gale62college}. An instance of the Stable Marriage problem
involves two sets of participants, conveniently called the set of men
$M$ and the set of women $W$. A common assumption is that every member
of each gender has strict preferences over the members of the opposite
gender\footnote{\label{foot:SMTI}The Stable Marriage with Ties problem
(SMT) raises issues that will not be addressed in this paper. For
existing studies on SMT, see
\cite{iwama99stable,manlove02structure}.}. The purpose of the
theory is to find and describe the stable matchings (or configurations) between $M$ and $W$. A matching $\mM$ is said to be unstable if there is a pair ($m$,$w$)
where each one prefers being matched with the other rather than being in its current situation in $\mM$.  
This pair is said to \emph{block} the matching $\mM$, and is called a blocking pair for $\mM$. A stable matching is a matching with no blocking pair. Using a concept of \emph{proposals}, Gale and Shapley have shown that all instances of the marriage problem possess at least one stable state that can be reached in $O(hf)$ proposals, where $h$ is the number of men and $f$ the number of women.

If there is only one set of participants (called \emph{peers}), where anybody can be matched with anybody, then we get a different problem, called the Roommate problem. This change has two important consequences. Firstly, the existence of a stable configuration is no longer guaranteed, and secondly, a proposals algorithm like the Gale-Shapley algorithm may not converge, even if a stable configuration exists.

The first issue has been addressed by Irving in \cite{irving85}. The Irving algorithm finds a stable configuration to the roommates problem, if there is one, or it indicates that no solution exists. 
However, to the best of our knowledge, an adaptation of the Gale-Shapley algorithm to the roommates problem is still to be developed. One of the contributions of this work is a natural extension of Gale-Shapley to the stable roommates theory for the special case of acyclic preferences (see Section \ref{sec:stability}).

Finally, the roommates problem can be further generalized by allowing  any peer $p$ to establish a number $b(p)$ of simultaneous partnerships (instead of a single one as in the classical roommate matching problem).
This generalization is often called many-to-many matching in the bipartite case or \emph{b}-matching in the general case\cite{fleiner03stable,konishi03credible}. Cechl\'arov\'a and Fleiner~\cite{fleiner05generalization} show how a \emph{b}-matching problem can be transformed into an equivalent 1-matching problem.
We propose to use \emph{b}-matching to model the connections in a P2P network, as 
detailed in the following section.

\section{Networks as matching instances}
\label{sec:model}

P2P networks are formed by establishing an overlay network between peers. Any peer acts both as a server and a client. Each peer $p$ uses a bounded number of connections. As the network evolves,
peers continuously seek after new (or better) partners. 
Each protocol implements its own algorithm for this searching phase. But in most of the cases, its output can be seen as a preference list over the contacts. Thus, a P2P protocol algorithm for connecting peers can be modeled as an instance of a \emph{b}-matching problem.

As for the stable roommates problem, we consider a set $P$ of $n$ peers. All possible connections between the peers are defined by \emph{an acceptance graph}. Each peer $p$ has a \emph{quota} $b(p)$ on the number of mates (connections). In addition, all neighbors of $p$ from the acceptance graph are sorted according to a given preference system and form a preference list denoted by $L(p)$. $L(p,q)$ denotes the position (rank) of peer $q$ in $p$'s list. In other terms, $L(p)$ is a permutation of all neighbors of $p$.
If $L(p,q) < L(p,r)$ then we say that $p$ \emph{prefers} $q$ to $r$. The best rank corresponds to $1$.
The \emph{degree} of a peer is the length of its preference list.
If we denote by $L$ a vector of preference lists corresponding to all peers, then $(P,L)$ (if there is no ambiguity $P$ is omitted)
 defines \emph{an instance}
of the Roommates Problem. 

For simplicity we consider only
\emph{undirected acceptance graphs}: $p\in L(q)$ iff $q\in L(p)$.
There is no loss of generality since pairs are formed only between peers that mutually accept  each other. Let $m$ denote the number of edges of the acceptance graph.

When a partnership is established between two peers $p$ and $q$, we say
that each one is a \emph{mate} of the other, or equivalently that the \emph{pair} $\{p,q\}$ is formed. A \emph{configuration} $C$ is defined as a set of formed \emph{pairs} $\{p,q\}$ such that each pair $p$ has at most $b(p)$ mates. Some peers may be single (i.e. not paired).
The set of all configurations is called
$\mc{C}$. It contains the trivial configuration $C_\emptyset$, where
no peer is paired.

In a configuration $C$, we say that $p$ is \emph{under-mated} if it has less than $b(p)$ mates in $C$. A \emph{blocking pair} for a configuration $C$ is a pair $\{p,q\}\notin
C$ such that each member of the blocking pair is either under-mated or 
prefers the other to its worst mate in $C$.

We propose to model the evolution of the system through \emph{initiatives}, a natural extension of the
Gale-Shapley initiative algorithm \cite{gale62college}. An initiative is  the process by which a peer may change its mates. Given a configuration $C$, we say that peer $p$ \emph{takes the initiative} when it proposes to other peers to be its new mate. Basically, $p$ may propose partnership to any acceptable peer. However, a new partnership is only interesting when a blocking pair exists.  If a peer $p$ is part of a blocking pair $(p,q)$ and elopes with $q$,
the initiative is then called \emph{active} because it modifies the configuration (both peers will change their set of mates). 

To find such a new mate, $p$ searches its preference list avoiding peers that do not improve  its situation.
We identify several strategies depending on how $p$ searches its preference list:
\begin{itemize}
\item \vspace{-2mm} \emph{best mate}:  $p$ seeks the best peer with which it forms a blocking pair.
\item \vspace{-2mm} \emph{decremental mate}:  $p$ circularly scans $L(p)$ starting from the position of the previous initiative.
\item \vspace{-2mm} \emph{random mate}:  $p$ chooses at random among the blocking pairs it belongs to. 
\end{itemize}
\vspace{-2mm}

Best mate seems to be the best strategy from a peer's point of view. However,  when making proposals takes a valuable time, it may not be realistic. For this reason, we consider the two other types of initiative which are more suited to model simultaneous asynchronous initiatives. 
For instance, consider an application where peers are not aware of their neighbors' value.
To try to find a better mate, a peer will select a random neighbor, probe it and keep collaborating
with it if it is more interesting than some previous mate.
This is the random mate strategy.

Let us now consider how preferences usually appear in P2P networks.

\section{Acyclic networks}
\label{sec:acyclic}

In this section, we show that most current P2P networks 
tend to conclude partnerships based on preferences which appear to be acyclic.
 A \emph{preference
  cycle} between $k\geq 3$ peers $p_1...p_k$ occurs if $p_i$ prefers
$p_{i+1}$ to $p_{i-1}$ (modulo $k$). A preference instance is
\emph{acyclic} if it contains no preference cycle.

First consider networks where mates are selected according to some
inherent capacity like available bandwidth\footnote{Network available bandwidth often mainly depends on the type of the peer Internet connection and how much is consumed by other concurrent applications.}, computing capacity, or storage capacity. 
In such a system, a peer $p$ possesses an
intrinsic mark $m(p)$ acknowledged by all the peers it knows.  Peers
with higher marks are preferred.  The preference lists resulting
from such a policy are called \emph{global preferences}.  Consider a
preference chain $p_1...p_k$ where peer $p_i$ prefers $p_{i+1}$ to
$p_{i-1}$. As marks increase along the chain, it cannot form a
preference cycle, and global preferences are always acyclic.

As an example, consider the ``Tit-for-Tat'' strategy of
BitTorrent~\cite{cohen03incentives}. Each peer prefers to exchange
with peers with the best upload capacity: as such peers provide data at a higher rate,
they appear as best uploaders as soon as a steady sequence of chunk
exchanges is initiated.  BitTorrent's Tit-for-Tat policy is thus close to a global
preference system according to upload capacity. In addition to best
uploaders, each peer also serves a random contact. This ``generous''
connection can be seen as a probing mechanism for finding mates with
better upload capacity.

However, it should be noted that peer selection also relies on
the complementarity of file chunks. This can be further modeled by a
second type of preference system that we call \emph{complementary
  preferences}.  In such a system, all peers try to get the same set
of resources (such as file chunks in cooperative file download).
Each peer then prefers to exchange resources with peers possessing
the largest number of its missing resources. It can be shown that such
complementary preferences are also acyclic.
Notice that this kind of preference changes as blocks are
downloaded. However, the peers with the largest complementary set of blocks
are those enabling the longest block exchange sessions.  In its most
general form, the selection of peers for cooperative file download
can be seen as a mix of two acyclic preference systems.

Finally, we identify a third class of acyclic preference systems
where each peer $p$ gives a mark $m(p,q)$ to each peer $q$ it knows
in such a manner that marks are symmetric: $m(p,q)=m(q,p)$ for all
$p,q$.  Each peer prefers to pair with peers with the best marks.
  Such a preference system is said to be \emph{symmetric}.  Again the marks increase
along a preference chain, preventing the existence of a preference
cycle. The simplest example of such a preference system comes from
latency optimization. Consider an overlay scheme such as
Pastry~\cite{pastry} that is optimized by selecting contacts with
the smallest round trip time (RTT) in the physical network. As the RTT
is a symmetrical measure (on average), it results in symmetric
marks.

In fact, any selection mechanism induced by proximity according to
some distance function results in such symmetric preferences.  For
example, massively multiplayer online games (MMOG) require to connect
players with nearby coordinates in a virtual
space. This can be modeled by
symmetric preferences based on distance in the virtual space.
Similarly, some authors propose to connect participants of a file
sharing system according to the similarity of their
interest~\cite{fessant04clustering,sripanidkulchai03efficient}.
Any such preference system based on proximity is symmetric
and thus acyclic.

We have seen that many P2P networks are formed through peer
selection algorithms that can be modeled by preference instances
that are acyclic. We now consider the stability properties %properties
of such preference systems.

\section{Stability result}
\label{sec:stability}

While it is difficult to find the stable solutions for general preferences in the roommates problem, the issue for acyclic preferences is much simpler, as shown by Theorem \ref{thm:acyclicunique}.
\begin{thm}
\label{thm:acyclicunique}
An acyclic $b$-matching preference instance always has a unique stable configuration.
\end{thm}
\begin{Proof}
We first prove by contradiction
that there can be at most one stable solution when preferences are acyclic. Suppose $A$ and $B$ are two distinct stable configurations of the instance. There exists a peer $p_1$ with different mates in $A$ and $B$. Let $p_2$ be the best mate among the mates $p_1$ is matched with in $A$ or $B$, but not in both configurations. Assume, without loss of generality, that $p_1$ is mated with $p_2$ in $A$, but not in $B$.
As $B$ is stable, $p_2$ has $b(p_2)$ mates in $B$ it prefers to $p_1$ (otherwise $\{p_1,p_2\}$ would be a blocking pair for $B$).  At least one of them is not its mate in $A$; let  $p_3$ be the best ranked.
For a similar reason, $p_3$ has a mate $p_4$ in $A$ and not in $B$ it prefers to $p_2$. We iterate the process to construct a sequence of peers $(p_i)_{i\geq 1}$. As the set of peers is finite, a peer $p_k$ is found that is already present in the sequence. Let us take the smallest $k$ such that $p_i=p_k$ for some $i<k$. The choice of $p_{i+1}$ implies that $p_i$ prefers $p_{i+1}$ to $p_{k-1}$. By construction, the
circular list $p_i,p_2,\ldots,p_{k-1}$ is a preference cycle. This contradicts the fact that the instance is acyclic.

To ensure the existence of a stable solution, we now prove that when preferences are acyclic, a sequence of active initiatives (i.e., initiatives that change the configuration) never goes twice through the same configuration. As there is a finite number of possible configurations, if we keep altering the configuration through initiatives, we eventually reach a configuration that cannot be altered with any initiative: a stable configuration.

The proof is simple. If a sequence of initiatives induces a cycle of at least two distinct configurations, then one can extract a preference cycle: let $p_1$ be a peer whose mates change through the cycle. Call $p_2$ the best peer $p_1$ is unstably paired with during the cycle, and  $p_3$ the best peer $p_2$ is unstably paired with during the cycle. $p_1$ is not $p_3$ and $p_2$ prefers $p_3$ to $p_1$, otherwise the pair $\{p_1,p_2\}$ would not break during the cycle. Iterating the process, we build a sequence of peer $(p_k)$ such that $p_k$ prefers $p_{k+1}$ to $p_{k-1}$, until we find $i<j$ such that $p_i=p_j$. The circular list $(p_i,p_{i+1},\ldots,p_{j-1})$ is a preference cycle.
\end{Proof}

Theorem~\ref{thm:acyclicunique} can be also proved using Tan's decomposition theory \cite{tan91necessary}. However, our proof has the advantage of leading to Corollary \ref{cor:acycl}.

\begin{cor}
\label{cor:acycl} Any sufficiently long sequence of active initiatives leads to the unique stable configuration.
\end{cor}

Any initiative algorithm, starting from any configuration, computes the stable configuration if we assume it gives enough initiatives to active peers. This statement alone gives no guarantee on the convergence speed (except a factorial bound derived from the number of possible configurations), but more insight is given next.

\section{Convergence results}
\label{sec:convergence}

In order to understand acyclic dynamics, we shall introduce the concept of loving pair. A loving pair $\{p,q\}$ is a pair of peers such that peer $p$ is ranked first by peer $q$ and \emph{vice versa}. Loving pairs are the key to understanding acyclic preferences and convergence to the stable state. They have two main properties: first, loving pairs are unbreakable. Once peers of a loving pair are matched together, no sequence of initiatives can unmatch them. The second property is given by Theorem~\ref{lemGai}.
An instance is said to be trivial when all preference lists are empty. 

\begin{thm}\label{lemGai}
Any non-trivial acyclic preference instance always has at least one loving pair.
\end{thm}

\begin{Proof} Consider a non-trivial preference instance.
There exist $2$ peers $p_0$, $p_1$ such that $L(p_0,p_1)=1$. If $L(p_1,p_0)=1$, then $\{p_0,p_1\}$ is a loving pair. Otherwise, there exists $p_2\neq p_0$ such that $L(p_1,p_2)=1$. If we continue this process, we eventually find a loving pair $p_i$, $p_{i+1}$ such that $L(p_i,p_{i+1})=L(p_{i+1},p_i)=1$.
If this is not the case, we construct a sequence $p_0,\ldots,p_i,\ldots$ such that $L(p_{i-1},p_i)=L(p_i,p_{i+1})=1$, with $p_{i-1}\neq p_{i+1}$. As the number of peers is finite, the sequence
loops, producing a preference cycle.
\end{Proof}

Because of loving pairs, the stable solution can be constructed pair by pair through initiatives. This is stated in Theorem~\ref{thm:optimal_sequence}.

\begin{thm}
\label{thm:optimal_sequence}
For any acyclic preferences instance, starting from any initial configuration $C$, there exists a sequence of at most $\frac{B}{2}$ initiatives leading to the stable solution, where $B=\sum_{p\in P}b(p)$.
\end{thm}

\begin{Proof}
Theorem~\ref{thm:acyclicunique} guarantees that a stable configuration
exists for any acyclic preference instance. We show that this stable state can be reached by matching loving pairs. We say that a 
pair is \emph{stable} when no sequence of initiatives can break it. All stable pairs $\{p,q\}$ can be taken out of consideration: we can virtually remove
them from the acceptance graph, erasing each peer from the preference list of the other,
and decrementing the quotas $b(p)$ and $b(q)$. 
Similarly, a peer $p$ that appears in $b(p)$
stable pairs will never change its mates,  %can be virtually removed from the instance.
which is equivalent to considering a preference instance where $p$ has been removed.
In both cases, we obtain a smaller, but strictly equivalent preference instance.
Moreover, this preference instance remains acyclic.

Starting from any configuration, we first remove all stable pairs
and peers with a full quota of stable pairs as described above.
The equivalent preference instance is acyclic.
As long as it is non-trivial, Theorem~\ref{lemGai}
implies the existence of loving pairs.
%As there must exist a loving pair, we 
We give a best mate initiative to one peer of a loving pair.
% in the current modified preference list. 
The loving pair is then formed and it is stable. We can remove it from consideration. It results in an equivalent preference instance where $B$ is decremented by 2. If one of the peers
has now a $0$ quota, it can be removed from the preference instance.
%removing $2$ connections from $B$.
By iterating this process at most $\frac{B}{2}$ times, we end up in a configuration with a trivial equivalent
preference instance 
%any peer $p$ has quota $b(p)=0$ or has an empty 
where all preference lists are empty. This final configuration is thus
the stable solution of the initial preference instance.
\end{Proof}

The above algorithm for computing an optimal initiative sequence is hard
to implement in a massively distributed environment. However the key of Theorem \ref{thm:optimal_sequence} is that the stable solution is made of at most $B/2$ stable matchings, and that at each moment, one of them is a loving pair for the current configuration. With the best mate strategy, 
a loving pair is formed as soon as one of its peers has the initiative.
%suffices one peer of a loving pair to take initiative to form it. 
We can estimate the convergence speed by estimating the time needed to match loving pairs. For instance, consider two simple random algorithms: periodic and Poisson.

In the periodic algorithm, each peer takes a best mate initiative every $t$ seconds. Due to network latencies, we consider that the order of initiative may be different in each period. By  Theorem~\ref{lemGai}, with \emph{best mate}, a loving pair is formed after at most $t$ seconds. Thus, in an acyclic preference instance, with periodic initiatives, the stable configuration is reached after at most $\frac{B}{2}t$ seconds.

In the Poisson algorithm, at any step a peer drawn uniformly at random with probability $\frac{1}{n}$ takes an initiative. A classical balls and bins result states
that, with high probability (w.h.p.), each peer will have taken the
initiative at least once after $n \log n$ drawings. Thus, in an acyclic
preference instance, with best mate Poisson initiatives, the stable
configuration is reached in $O(nB \log n)$ initiatives w.h.p.. 

The mean convergence time is much smaller. Considering that in each unstable configuration there exists at least $2$ peers from a loving pair, the mean time between the creation of two stable pairs in a best mate Poisson initiatives sequence is at most $\frac{n}{2}$. As a consequence, for any acyclic instance and any starting configuration, the mean time to reach the stable state with best mate Poisson initiatives is at most $\frac{nB}{4}$.

\section{Future Work and Conclusions}
\label{sec:future}

In the present work we have given upper bound estimates of the convergence time. These results are based on the existence of at least one loving pair at every step.
Preliminary simulation results lead us to believe that the $\frac{nB}{4}$ bound on is tight for global preferences, when $b=1$ and the acceptance graph is complete. In this case the only loving pair is composed of the two best globally ranked peers, which do not yet have paired together. However, we suspect
that several other P2P networks preference systems, such as symmetric preferences, 
may contain a large number of loving pairs
at a time. These systems should converge much faster. We plan to identify and analyze such preference systems.

Note, that our convergence results assume that the preference lists are static.
However, for most P2P networks, the set of peers and their preference lists evolve in time. Further work will consider the impact of such dynamics on the stable
configuration of the system. A major interest is to compare convergence speed
to the system evolution speed. As long as it is fast enough, we can expect 
that the system will smoothly follow the evolving stable configuration target.

We have shown how collaboration selection algorithms can be modeled using 
the matching theory. In fact, most of these algorithms lead to acyclic preference instances and we have proved that such preference systems always evolve towards a stable configuration. Additional insight was given on the convergence time which is required to reach this stable configuration.

\textsl{\textbf{Acknowledgment:} The authors wish to thank James Roberts for his helpful comments} 

%\vspace{-4mm}
\bibliographystyle{alpha}
\bibliography{maria}

\begin{thebibliography}{IMMM99}

\bibitem[CF05]{fleiner05generalization}
K.~Cechl\'arov\'a and T.~Fleiner.
\newblock On a generalization of the stable roommates problem.
\newblock {\em ACM Trans. Algorithms}, 1(1):143--156, 2005.

\bibitem[Coh03]{cohen03incentives}
B.~Cohen.
\newblock Incentives build robustness in bittorrent.
\newblock In {\em P2PECON}, 2003.

\bibitem[FHKM04]{fessant04clustering}
F.~Fessant, S.~Handurukande, A.~Kermarrec, and L.~Massoulie.
\newblock Clustering in peer-to-peer file sharing workloads.
\newblock In {\em Proceedings of the 3rd International Workshop on Peer-to-Peer
  Systems (IPTPS)}, 2004.

\bibitem[Fle03]{fleiner03stable}
T.~Fleiner.
\newblock The stable b-matching polytope.
\newblock {\em Mathematical Social Science}, 46(2):149--158, 2003.

\bibitem[GS62]{gale62college}
D.~Gale and L.S. Shapley.
\newblock College admissions and the stability of marriage.
\newblock {\em American Mathematical Monthly}, 69:9--15, 1962.

\bibitem[IMMM99]{iwama99stable}
K.~Iwama, S.~Miyazaki, D.~Manlove, and Y.~Morita.
\newblock Stable marriage with incomplete lists and ties.
\newblock In {\em ICALP}, pages 443--452, 1999.

\bibitem[Irv85]{irving85}
R.~Irving.
\newblock An efficient algorithm for the stable roommates problem.
\newblock {\em J. of Algorithms}, 6:577--595, 1985.

\bibitem[KU03]{konishi03credible}
H.~Konishi and M.~Unver.
\newblock Credible group-stability in many-to-many matching problems.
\newblock Technical Report 570, Boston College Department of Economics,
  September 2003.

\bibitem[Man02]{manlove02structure}
D.~Manlove.
\newblock The structure of stable marriage with indifference.
\newblock {\em Dis. Appl. Math.}, 122:167--181, 2002.

\bibitem[RD01]{pastry}
A.~Rowstron and P.~Druschel.
\newblock Pastry: Scalable, decentralized object location, and routing for
  large-scale peer-to-peer systems.
\newblock {\em Lecture Notes in Computer Science}, 2218:329--??, 2001.

\bibitem[SMZ03]{sripanidkulchai03efficient}
K.~Sripanidkulchai, B.~Maggs, and H.~Zhang.
\newblock Efficient content location using interest-based locality in
  peer-to-peer systems.
\newblock In {\em INFOCOM}, 2003.

\bibitem[Tan91]{tan91necessary}
Jimmy J.~M. Tan.
\newblock A necessary and sufficient condition for the existence of a complete
  stable matching.
\newblock {\em J. Algorithms}, 12(1):154--178, 1991.

\end{thebibliography}

\end{document}